\documentclass{article} \usepackage{amsmath} \usepackage{amssymb}
\usepackage{epsfig}
\usepackage{amscd}
\newtheorem{theorem}{Theorem}

\newcommand{\cD}{\mathcal{D}}

\newcommand{\cM}{\mathcal{M}}
\newcommand{\cH}{\mathcal{H}}

\newcommand{\cL}{\mathcal{L}}

\newcommand{\Om}{\Omega}
\newcommand{\Si}{\Sigma}
\newcommand{\La}{\Lambda}
\newcommand{\rme}{\textrm{e}}
\newcommand{\rmO}{\textrm{O}}

\newcommand{\eg}{\textit{e.g.}}
\newcommand{\ie}{\textit{i.e.}}

\newcommand{\etc}{\textit{etc.}}

\newcommand{\be}{\begin{equation}}
\newcommand{\ee}{\end{equation}}
\newcommand{\ba}{\begin{align}}
\newcommand{\ea}{\end{align}}
\newcommand{\ra}{\rightarrow}

\font\openface=msbm10 at10pt
\def\Minkowski     {{\hbox{\openface M}}}

\def\Reals         {{\hbox{\openface R}}}

\def\Euclid        {{\hbox{\openface E}}}

\def\Lor  {\cL_0}
\def\Aut {\mathop{\rm Aut} \nolimits}	 

\begin{document}
\title {Discreteness without symmetry breaking: a theorem}
\author{
Luca Bombelli\footnote{Department of Physics and Astronomy,
  University of Mississippi, University, MS 38677, USA},
Joe Henson\footnote{Institute for Theoretical Physics,
  University of Utrecht, Minnaert Building, Leuvenlaan 4,
  3584 CE Utrecht, The Netherlands.}
\ and Rafael D. Sorkin\footnote
{Department of Physics, Syracuse University, Syracuse NY 13244-1130, USA
and Perimeter Institute, Waterloo, N2J 2W9, Canada.}
}
\date{April 30, 2006} 
\maketitle

\begin{abstract}
\noindent
This paper concerns sprinklings into Minkowski space (Poisson
processes).  It proves that there exists no equivariant measurable
map from sprinklings to spacetime directions (even locally).  Therefore,
if a discrete structure is associated to a sprinkling in an intrinsic
manner, then the structure will not pick out a preferred frame, locally
or globally.  This implies that the discreteness of a sprinkled causal set
will not give rise to ``Lorentz breaking'' effects like modified
dispersion relations.  Another consequence is that there is no way to
associate a finite-valency graph to a sprinkling consistently with
Lorentz invariance.

\end{abstract}
\vskip 1cm

\section{Introduction}

Most approaches to quantum gravity aspire to replace the continuous
spacetime of general relativity by something more fundamental,
often something discrete.
The continuum can survive in such an approach only as an ``emergent''
description of the more fundamental structure, and so the
question arises of how to set up the correspondence
between the discrete structure and the approximating spacetime
(or space, in the case of canonical quantum gravity).
Appealing to the ideas of statistical geometry is one possibility
for doing this.
Indeed,
this method has for some time been central to the causal set program
\cite{SpacetimeAsCS,Sorkin:2003bx,henson:2006aa}
and has more recently been applied to loop quantum gravity as well
\cite{Bombelli:2004si}.
In the case of causal sets, a statistical correspondence is
the only kind that has ever been found to be consistent with the basic
postulate that spacetime volume reflects number of causet elements.
It is then an inevitable (and presumably welcome) consequence that
Lorentz invariance is recovered in the effective continuum description.
(In this sense, causal set theory {\it must} respect Lorentz invariance.)
The case of loop quantum gravity is similar even though one is dealing
with space rather than spacetime.
Because of the positive signature of the metric, a ``regular lattice''
(\eg, cubic) can now reproduce volume adequately.
However, if one wishes to recover surface area too, then --- once again
--- a statistical correspondence is the only known possibility.
(In this sense, loop quantum gravity must respect rotation invariance.)

Let us assume that our discrete structure is a set of
elements (or ``vertices'')
endowed with some extra information (an order relation, a
labelling by spins, \etc).
Let there also be a method of
inducing one of these discrete structures on each locally finite
subset of spacetime (or space),
\ie, a map $C$ from the space of such sets of points
of the continuum to the space of discrete structures.
If the fundamental
structure were a causal set (a locally finite partial order), this
map would associate to each set of points drawn from a Lorentzian manifold
the order induced on them by the causal order of the
manifold.\footnote
{Although this is the most obvious possibility for $C$, one has also
  considered generalizations in which the map $C$ itself contains a
  random element.  We would not expect such generalization to alter the
  main conclusions of this paper.}
If the structure were a graph playing the role of a spatial
configuration, the map would associate to each set of points drawn from
a Riemannian manifold a set of vertices, identified with the points
themselves or an appropriately defined dual set, and a set of 
edges,
obtained for example as a result of applying the delaunay or 
Voronoi
construction, respectively.\footnote
{Such constructions cover most of the approaches to quantum gravity
  that are popular today, but they do not represent the most general case.
  For example, a covering of a manifold by open sets would not give rise
  to a discrete structure whose vertices corresponded to specific points
  in the manifold.}

A discrete/continuum correspondence is needed to let us
connect properties of the new fundamental structure with known
physics.
Given a discrete structure $\cD$
and a candidate approximating manifold $\cM$,
the methods of statistical
geometry establish criteria for answering the question
``Is $\cM$ a good approximation to $\cD$?''.
(We will write $\cM\approx\cD$ for this.)
In any manifold possessing a volume measure, we can choose a locally
finite, uniformly random set of points by the {\it Poisson process}
in which the probability of finding $n$ points in a region of
volume $V$ is
\be
\label{e::poisson}
P(n)= \frac{(\rho V)^n\, \rme^{-\rho V}}{n!}\;,
\ee
where $\rho$ is some fundamental density; we call such a set of
points a ``sprinkling".  Let us now induce our discrete structure
on the sprinkled points, and notice that the map from sets of
points in $\cM$ to discrete structures is
in general
many-to-one.
We then want to
say that $\cM\approx\cD$ iff $\cD$ arises ``with high probability''
from a sprinkling of $\cM$.
Conversely, it we begin with $\cM$ and derive a $\cD$ from it by
sprinkling, then we will almost always obtain  a $\cD$ to which
$\cM$ is a good approximation.\footnote
{The underlying idea is that $\cM\approx\cD$ iff $\cD$ is a ``typical
  result'' of sprinkling $\cM$.  When $\cM$ is compact (with boundary)
  one can try to make the idea of typicality precise by identifying it
  with the quoted requirement of ``high probability''.  For $\cM$ of
  infinite volume (\eg, $\cM=\Minkowski^n$) this idea evidently won't work
  as such; for the probability to obtain any given $\cD$ is clearly
  zero.  Nevertheless, any attribute of $\cD$ that is present with
  probability 1 (relative to the Poisson process in $\cM$) can be taken
  to obtain whenever  $\cM\approx\cD$.}
As we will see, this way of setting
the discrete/continuum correspondence helps to
preserve the well-observed symmetries
of the continuum.  (And, as remarked above, it is the only known way,
consistent with the indefinite signature of the spacetime metric, to
implement the assumption that number equals volume.)

If the set of all discrete structures  $\cD$  is to be a ``history-space'',
rather than a space of possible spatial configurations, it is
reasonable to require that  at least some of the  $\cD$  should
admit Minkowski
space as a good approximation (at some sufficiently large scale,
in the same way as some paths
of the non-relativistic path-integral track classical
trajectories at large scales, even though they are fractal
on short scales --- such paths dominate the integral in
semi-classical situations).  It is therefore interesting to ask how
well the symmetries of Minkowski space can be preserved in the emergent
continuum.

In what sense can we expect our discrete structure to be
Lorentz invariant --- what sense of the term is physically
relevant here?  An answer was proposed in Ref.\ \cite{Dowker:2003hb}:
that the discreteness must not, in and of itself, serve to pick out a
preferred frame (or frames) of reference.
This resembles one of the definitions of Lorentz invariance
(that ``the laws of nature'' not serve to pick out a frame),
and it is plausibly the criterion most relevant for
phenomenology.\footnote
{We do not mean to imply that it is the only criterion of interest.  For
  example, translation invariance in this sense does not, in and of
  itself, imply that some notion of momentum will be defined and obey a
  conservation law.}
In Ref.\ \cite{Dowker:2003hb} we presented strong evidence that causets
produced by sprinkling into Minkowski spacetime meet this criterion, but
a skeptic could still have found grounds for doubt.  In this paper, we
prove a theorem that we believe removes most of the remaining doubt.

The fact that the process of ``causet sprinkling'' in Minkowski space is
Lorentz invariant is an important first step in the argument.
(In this process we include both the Poisson sprinkling
as such and the subsequent induction of the causal order.  Both
steps are manifestly Lorentz invariant since they depend only on the
volume element and the causal structure of the spacetime,
respectively).
But Lorentz invariance of the
resulting causal set in the above sense does not immediately
follow.  Consider by analogy a game of fortune in which a circular
wheel is spun to a random orientation.  While the {\it distribution}
of final directions is indeed rotationally invariant, a {\it particular
outcome} of the process is certainly not.  (A form of ``spontaneous
symmetry breaking'', perhaps).
Likewise, a
particular outcome of the Poisson process might be able to prefer a
frame, even though the process itself does not.

So, the question becomes: Is it possible to use a sprinkling of
Minkowski space to select a preferred frame?  We will prove a theorem
that answers ``no'' to this question.  In fact, it answers the slightly
more general question whether a sprinkling can pick out a preferred
time-direction (which is certainly possible if an entire frame can be
derived.)  Below, we formalise the notion of deriving a direction from a
sprinkling, and we prove a theorem showing that this cannot be done.  In
this sense, the situation with sprinklings of Minkowski space is even
more comfortable than that with sprinklings of Euclidean space.  It is
possible to associate a direction from the rotation group to a point in
such a sprinkling, as discussed later (although this will not stop
anyone from maintaining that a gas behaves isotropically in the
continuum approximation; these locally defined directions have little
significance at that level), but the non-compactness of the Lorentz
group makes the
Lorentzian case different.

Based on the theorem, we can assert the following.  Not only is the
Poisson process in Minkowski space Lorentz invariant, but the
\textit{individual realizations} of the process are also Lorentz
invariant in a definite and physically important sense.

Another question one might ask is: how easy is it to come up with maps
taking sprinklings to discrete structures in general?  Is this feature
of causal sets unusual, or is it generic?  Our theorem is relevant to
this question as well.  As a corollary, it implies that no finite
valency graph can be associated to a sprinkling of Minkowski space
consistently with Lorentz invariance.  This rules out the use of the
sprinkling technique to find Lorentz invariant spin-foams or
relativistic spin-lattices.
Precisely this problem was encountered by T.D. Lee's ``random lattices'' in
Minkowski space.

\section{A theorem}

Let $\Minkowski^n$ be $n$-dimensional Minkowski space,
and let $\Lor$ be the connected component of the identity in
$\rmO(n-1,1)$, the full Lorentz group of $\Minkowski^n$.
We will call $\Lor$ simply ``the Lorentz group''.
Fix a point $O$ (``the origin'') in $\Minkowski^n$ and let $\Lor$ act on
$\Minkowski^n$ with $O$ as its fixed point.

An individual realization of a Poisson process in $\Minkowski^n$ is
almost surely a locally finite subset of $\Minkowski^n$ (\ie, a
collection of point-events of $\Minkowski^n$ with no accumulation point
anywhere).  The space of all such subsets or ``possible sprinklings'',
we will denote by $\Om$.
A Poisson process is captured mathematically by a probability
measure $\mu$ on $\Om$.
Formally, it is
the stochastic process defined by the triple $(\Om, \Si, \mu)$,
where
$\mu: \Si \ra \Reals$
and $\Si$ is
the $\sigma$-algebra of all measurable subsets of $\Om$,
as defined, for example, in Ref.\ \cite{Stoyan:1995}.

Clearly the action of $\Lor$ on spacetime points induces an action on
collections of spacetime points.  In particular it induces an action
on $\Om$ under which $\Si$ is left invariant.
It is known \cite{Stoyan:1995} that a Poisson process in
$\Minkowski^n$ is invariant against any volume-preserving, linear 
transformation
of $\Minkowski^n$,
the ultimate reason being that Eq.\ (\ref{e::poisson}) only refers to volumes.
Consequently it is invariant under the action of $\Lor$; that is, the
probability of a (measurable) set of possible sprinklings is the same as
that of the set obtained applying a Lorentz transformation to it:
\be
   \label{e::LI}
   \mu = \mu \circ \La  \;, \quad  \forall \, \Lambda \in \Lor \;.
\ee


Since we want to prove a local theorem and not just a global one, let us
consider the existence of a preferred direction relative to a selected
point of $\Minkowski^n$, which we can take to be the origin $O$.
(In cases of interest, $O$ will actually be a point of the sprinkling,
but we don't need to assume that for purposes of the proof.)
The assertion that every sprinkling determines
a preferred direction at $O$ states at a formal level that there exists
a map $D$ from $\Om$ to $\cH$,
the hyperboloid of unit future-timelike vectors.
This map $D$ is the ``rule'' by which each individual set $\omega$ of
sprinkled points gives rise to the direction $D(\omega)$ at $O$.
Not every function   $D:\Om\to\cH$ is a valid candidate, however,
because we want the direction chosen by $D$
``to have come from the sprinkling and nothing else''.
As an example of what we {\it don't} want,
consider the map $D_X$ that takes every sprinkling $\omega$ to the same
vector $X$.  The ``distinguished direction'' defined by this map (namely
$X$) was clearly put in by hand; it has nothing to do with the sprinkling
from which it supposedly came.
%
%
In order to eliminate such specious rules, we should require
{\it equivariance} with respect to $\Lor$, \ie, we should require that
a Lorentz transformed sprinkling $\La\,\omega$ give rise to the
correspondingly transformed direction $\La\,X$.
In other words, we should require that the process $D$ of deducing a direction
from a sprinkling commute with Lorentz transformations (which is a
special case of the more general requirement, which we could impose with
equal justice, that $D$ commute with all of $\Aut{\Minkowski^n}$).
The equivariance of $D$ can be expressed by a commutative diagram:
\begin{equation*}
   \begin{CD}
     \Om    @> \La >> \Om    \\
     @VVDV            @VVDV  \\
     \cH    @> \La >> \cH
   \end{CD}
\end{equation*}


\begin{theorem}
In dimensions $n>1$
there exists no measurable equivariant map $D:\Om \ra \cH$,
\ie, there exists no measurable $D$ such that
\be
  \label{e::commute}
  D \circ \La = \La \circ D\;,
  \quad \forall \, \Lambda \in  \Lor \;.
\ee
\end{theorem}
\paragraph{Proof:}

Suppose that such a map $D$ exists.   Its
inverse $D^{-1}$ yields a well-defined map from subsets of $\cH$ to
subsets of $\Om$, and this in turn lets us define
a probability distribution
$\mu_D$ on $\cH$, as follows.
Since $D$ is measurable, it follows  by
definition  that the inverse image of each measurable subset of $\cH$ is
measurable in $\Om$, and we set
\ba
  \label{e::defmud}
  \mu_D:=\mu \circ D^{-1}.
\end{align}
Eq.\ (\ref{e::commute}) then implies that
$\La \circ D^{-1} = D^{-1} \circ \La$,
and using Eqs.\ (\ref{e::defmud}), (\ref{e::LI}), and
(\ref{e::commute}), respectively, we can see that
\ba
\mu_D &= \mu \circ D^{-1}
  = \mu \circ \La \circ D^{-1}  \nonumber\\
  &= \mu \circ D^{-1} \circ \La
  = \mu_D \circ \La\;, \qquad \forall\,\La \in \Lor \;.
\end{align}
This means that $\mu_D$ is a probability measure on the unit
hyperboloid $\cH$ that is invariant under
the action of $\Lor$.
But the non-compactness of the hyperboloid, related to that of
the Lorentz group, forbids this.  To see why, consider an open
set $U \subset \cH$ of compact closure; such a set is measurable
and has a finite measure $\mu_D(U)$.
We can assume without loss of generality that $\mu_D(U)>0$.
Now apply a boost $\La$
such that $U$ and its images $U_n:= \La^n\,U$ are all disjoint.
By Lorentz invariance, $\mu_D(U_n) = \mu_D(U)$, and therefore
by the additivity of the measure, $\mu_D(\cup_{i=1}^nU_i) = n\,\mu_D(U)$
for any $n$.  But this is impossible if $\mu$ is a probability
measure, because for $n$ sufficiently large,
$n\,\mu_D(U)$ would exceed any pre-assigned value,
which is absurd since probabilities cannot exceed unity.
Therefore no measurable $D$ obeying Eq.\ (\ref{e::commute}) can
exist.\footnote
{Notice that the measurability requirement on $D$ is extremely weak; we
  know of no use of non-measurable maps in physics, and in fact no such
  map can be explicitly specified.}
\hfill$\Box$

\paragraph{}
The theorem extends easily to other, similar statements about the
impossibility of sprinklings in Minkowski space breaking Lorentz
invariance:

\vskip 8pt
\noindent
(1) No {\it partially defined} equivariant measurable map from
sprinklings to directions can exist if its domain has nonzero measure.
(This rules out the possibility that a nontrivial subset of the
sprinklings could break Lorentz invariance, even if not all of them could.)

\vskip 8pt
\noindent
(2) A sprinkling cannot determine a timelike direction globally, in
Minkowski space as a whole (or in any Lorentz invariant subset of
Minkowski space).  Otherwise that direction could be defined as the
preferred one at each point, contradicting the result just proved.

\vskip 8pt
\noindent
(3) A sprinkling cannot determine a preferred {\it spacelike} direction at
any point of $\Minkowski^n$.
The proof proceeds like
the one for the timelike case, with $\cH$ replaced by the hyperboloid
of unit spacelike vectors at the origin.

\vskip 8pt
\noindent
(4) No finite set of timelike and/or spacelike directions at a
point (this includes a reference frame) can be associated to a
sprinkling consistently with Lorentz invariance.  In this case,
the map $D$ would go from $\Om$ to a product of various copies of the two
hyperboloids, possibly quotiented by the action of a permutation
group.

\vskip 8pt
\noindent
(5) A sprinkling cannot determine a preferred {\it location} in
Minkowski spacetime.  In other words, a Poisson sprinkled set of points
is as homogeneous as it is isotropic.  (Thus, our results hold for
the Poincar{\'e} group as well as the Lorentz group.)

\vskip 8pt
In each of these cases, there is no probability-measure on the
corresponding space that is invariant under the Lorentz (or 
Poincar{\'e}) group.
However, this is not true for infinite subsets of $\cH$, and a
countably
infinite set of directions \textit{can} be equivariantly associated
to a sprinkling.  An example is the set of directions from the
origin to all the sprinkled points in its future.  This is why
the theorem does not exclude the possibility of consistently
associating a causal set to a sprinkling.

If a finite set of directions cannot be associated to a
point in
a sprinkling of Minkowski space, consistently with Lorentz
invariance, it is clear that any method used to associate a
finite valency graph to such a sprinkling would violate Lorentz
invariance, regardless of any other properties the discrete
structure might have (such as labels on the edges or vertices, or the
presence
of higher-dimensional cells).
Some preferred direction would have
to be introduced before one could fill in edges between the points
of the sprinkling.

\subsection{An example}

The theorem is rather abstract and might conflict with the intuition of
people who are accustomed to working with Lorentz-violating
discretisations.  As an aid to intuition, we give an example in which
the construction of a direction-map $D$ fails.

In sprinklings of flat Euclidean space $\Euclid^n$,
it {\it is} possible to associate a direction to a point
in an equivariant (rotationally covariant) way.
An example is the map from a point in $\Euclid^n$ (possibly
one belonging to the sprinkling $\omega$) to the direction towards
the nearest sprinkled point.
This map {\it is} equivariant, since
rotating $\omega$ around our chosen
point and then finding the direction, gives the same result as
finding the direction and then rotating it.\footnote
{Whether a direction can be associated to the sprinkling as a whole is
  not clear to us, although one might expect that the answer is no.}

So, why does the analogous construction fail in the Lorentzian case?  The
answer is that there is no nearest neighbour to a given point in
a sprinkling of $\Minkowski^4$.  To put it another way, there is no
lower bound on the distance from the origin to a sprinkled
point.  Consider a point sprinkled at Lorentzian distance
$d$ from the origin.  The region within distance $d$ of the
origin is of infinite volume, extending all the way up the
light-cone, and we see from Eq.\ (\ref{e::poisson}) that there
are one or more points sprinkled into this volume with
probability 1.  Therefore, no matter how close a sprinkled point
is to the origin, there is always another point sprinkled closer.

\section{Conclusion}

As an example of how the theorem proved above can carry phenomenological
implications, consider a model of scalar field propagation in which we
discretise Minkowski space by replacing it with a sprinkled causal set,
as in Ref.\ \cite{dowker:2005}.  The theorem shows that no special frame or
direction $u^a$ can be picked out with respect to which one could
introduce Lorentz violating effects, even a $u^a$ that varies
stochastically with position.

The theorem also shows that a finite valency graph like a
``spin-foam'' cannot be
equivariantly associated to a sprinkling of full Minkowski space.
If we wanted our discrete structure to contain such a graph, we could
attempt to use some other continuum approximation scheme.
The problem is that no other way is known that preserves Lorentz
symmetry, when Minkowski space is to be the effective continuum
description.
%
%
Rather the two requirements of Lorentz invariance and discreteness seem
to lead to an unavoidable randomness already at the kinematic level:
a random discrete/continuum correspondence.
The causal set is a simple example of a structure amenable to this sort
of correspondence, and in fact the only example so far proposed in the
literature.
Others could be imagined (\eg, adding
distance information to the relations in the causal set), but we would
conjecture that
the causal set is, in some sense, the minimal Lorentz invariant discrete
structure from which the continuum can be reconstructed at macroscopic
scales.
Other such structures would then be expected to over-specify the
information needed to reconstruct the continuum.

These results apply, strictly speaking, only to full Minkowski space,
which of course is not realistic physically.
What of other
Lorentzian manifolds, \eg, large but finite regions of Minkowski?
In this case there would exist preferred directions that in principle
could be used to introduce graph-like structures.  This would be so even
in the continuum, because the boundary of the region would contain
directional information.  Indeed, we would conjecture that the boundary
would always enter essentially into any such scheme, rendering the
resulting phenomenological theory radically nonlocal.

The authors are grateful to the organisers of the ``Quantum
gravity in the Americas" conference at Perimeter Institute, where
this work was discussed.  JH was supported by DARPA grant
F49620-02-C-0010R at UCSD, where some work on the article was
carried out.
The work of RDS was partly supported by NSF grant PHY-0404646.

\bibliographystyle{h-physrev3}
\bibliography{refs}
\end{document}